\documentclass[aps,prl,preprint,superscriptaddress]{revtex4}
\usepackage{graphicx}
\usepackage{amsmath}
\usepackage[latin1]{inputenc}

\begin{document}


\title{Transport Detection of Quantum Hall Fluctuations in Graphene}



\author{Simon Branchaud}
\affiliation{Institute for Microstructural Sciences, National Research Council of Canada, Ottawa, ON, K1A 0R6}
\affiliation{Département de Physique, Université de Sherbrooke, Sherbrooke, QC, J1K 2R1}
\author{Alicia Kam}
\affiliation{Institute for Microstructural Sciences, National Research Council of Canada, Ottawa, ON, K1A 0R6}
\author{Piotr Zawadzki}
\affiliation{Institute for Microstructural Sciences, National Research Council of Canada, Ottawa, ON, K1A 0R6}
\author{François M. Peeters}
\affiliation{Department of Physics, University of Antwerp, 
Groenenborgerlaan 171, B-2020 Antwerp, Belgium}
\author{Andrew S. Sachrajda}
\affiliation{Institute for Microstructural Sciences, National Research Council of Canada, Ottawa, ON, K1A 0R6}

\date{\today}

\begin{abstract}
	Low temperature magnetoconductance measurements were made in the vicinity of the charge neutrality point. Two origins for the fluctuations were identified close to the CNP. At very low magnetic fields there exist only mesoscopic magneto-conductance quantum interference features which develop rapidly as a function of density. At slightly higher fields ($>$ 0.5T), close to the CNP, additional fluctuations track the quantum Hall sequence expected for monolayer graphene. These additional features are attributed to effects of locally charging individual quantum Hall (QH) localized states. These effects reveal a precursor to the quantum Hall effect (QHE) since, unlike previous transport observations of QH dots charging effects, they occur in the absence of quantum Hall plateaus or Shubnikov- de Haas (SdH) oscillations. From our transport data we are able to extract parameters that characterize the inhomogeneities in our device.
\end{abstract}

\pacs{}

\maketitle


	Since the first experimental realization of mechanically exfoliated graphene \cite{firsta} there has been growing interest in its material properties close to the charge neutrality point (CNP) \cite{rise, elecprop, Ishigami}. A finite resistance at the CNP provided strong evidence of the existence of electron and hole puddles which was confirmed directly by scanning probe techniques \cite{Martingraphenepuddles}.

	One of many remarkable properties of graphene in comparison with other 2DEGs is that in spite of its dramatically lower mobility, mesoscopic conductance fluctuations persist to surprisingly high temperatures \cite{Rothtdep, WLsuppr, Berger} as predicted \cite{rycerz}. On the other hand, effects which require time reversal symmetry in addition to phase coherence, such as weak localization, are suppressed in the absence of mechanisms enabling inter-valley scattering \cite{Savchenkomeso, WL, MccannWL}. Analysing low field magneto-transport measurements has become a critical tool to extract important scattering parameters. To make interpretation straightforward measurements are typically restricted to magnetic fields lower than those at which QHE features are visible. Graphene samples made via mechanically exfoliated techniques, however, tend to involve voltage contacts placed together at distances roughly equivalent to inhomogeneity length scales thus allowing the conductance to be sensitive to mesoscopic phenomena. In addition the spread in densities is significant in Graphene. This raises the question as to what low field measurements of magneto-conductance fluctuations in graphene are actually probing. In this paper, we find that the charging of individual QH localized states can dominate the magneto-conductance in transport experiments down to magnetic fields four times lower than any other QHE related feature such as QHE plateaux or SdH oscillations.

\begin{figure}[h]
 \includegraphics[width=55mm]{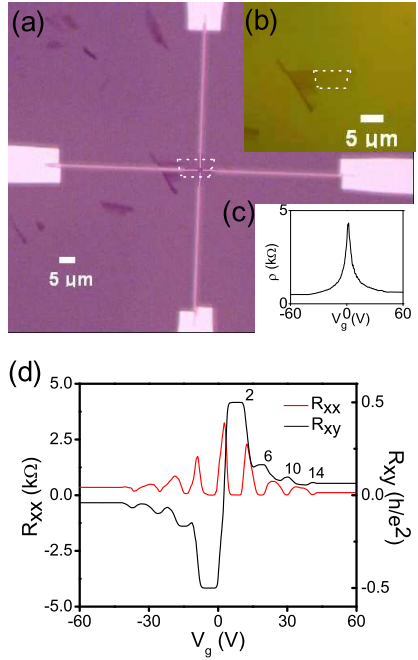}%
 \caption{(Color online) a. Picture of the graphene flake with deposited leads, notice the fine leads making a cross pattern in the very center.  The distance between leads at their closest point is of the order of 1 $\mu$m. b. Optical image of graphene flake, with the monolayer region outlined in white.  c. Resistance against back gate voltage at 4.2 K.  d. Longitudinal and transverse resistivity versus back gate voltage at 9 T, showing quantum Hall plateaus at filling factors $\nu = 4n-2$.}
 \end{figure}

	The graphene flake was mechanically exfoliated over an n-doped Si substrate covered with 300 nm of SiO$_2$. Figure 1(b) shows a micrograph of the graphene flake before and after the four Ti/Au contacts were deposited in a cross pattern. The spacing of the contacts at the closest point was ~ 1 $\mu$m. The carrier density was controlled by applying a voltage to the silicon back gate (a change of 1 V corresponds to a change in density of $7.1$$\times$$10^{10}$ cm$^{-2}$). Standard low noise AC transport techniques were used to make the measurements. One of the four contacts briefly overlapped a bilayer. To eliminate any possibility that this influenced the results, the measurements were repeated with this particular contact used both as voltage and current contact. In each case qualitatively identical results were obtained. Figure 1(c) shows the Dirac curve obtained using the Van der Pauw technique. Without any annealing the CNP occurred at +0.75 V. This small deviation suggests the absence of a significant incidental charged impurity local environment. To confirm the monolayer nature of the flake the Quantum Hall effect was measured. This is shown in Figure 1(d). The well established graphene monolayer sequence $\frac{h}{(4n+2)\cdot e^2}$ is observed. Magnetic field sweeps taken at high constant $1.96$$\times$$10^{12}$ cm$^{-2}$ electron density revealed SdH features down to the $h/22e^2$ (i.e. $n=5$) plateau.  From these characterizations we estimate the carrier mobility to be 19,000 cm$^2/$V$\cdot$s close to the CNP and 5300 cm$^2/$V$\cdot$s (5150 cm$^2/$V$\cdot$s) at a electron (hole) density of $7.3$$\times$$10^{11}$ cm$^{-2}$ We note, however, that no QHE feature was visible at any densities below 2T.

Conductance fluctuations measurements taken close to the CNP at low magnetic fields were visible with a root mean square exponential dependence up to 60 K, as shown in Figure 2. The detailed pattern of fluctuations was altered on cycling the temperature between 4 K and 100 K. Such properties are consistent with a quantum interference origin for the fluctuations. We also observe a broad resistance dip in magnetic field sweeps close to the CNP (disappearing at a density of $10^{11}$ cm$^{-2}$). This dip remained even at 100K confirming that it was not related to a broad weak anti-weak localization. Evidence for such an anti-weak localization feature in addition to a weak localization peak was obtained by subtracting the low temperature curves from the 100K data. A resistance dip which persisted to higher temperature has been previously reported in the literature and modelling found it to be consistent with the existence of electron and hole puddles \cite{Fuhrer}. This is consistent with our data as we were able to obtain an excellent fit to our dip using equation [3] from \cite{Fuhrer} if we used slightly different parameters for each field direction (attributed to our contact geometry).  The equation is based on an exact two fluid model \cite{guttalstroud} with an additional phenomenological conductivity term. 
\begin{equation}
\rho_{xx}(B) = \left(\sigma_{xx,1} + \frac{\sigma_{xx,0}}{[1 + (\mu B)^2]^{1/2}}\right)^{-1}
\end{equation}
with parameters for the left (right) side of the curve being $\sigma_{xx,1}$ = 2.68 (1.90) $e^2/h$, $\sigma_{xx,0}$ = 6.17 (6.95) $e^2/h$ and $\mu$ = 2.63 (2.93) m$^2/$V$\cdot$s.  The fit over the 100K trace is shown in Figure 2. Although the equation is strictly only valid at the CNP we find phenomenologically that the density range of the dip ($2$$\times$$10^{11}$ cm$^{-2}$) is consistent with the range of the puddle regime (see below) as obtained from the QH dot charging events.

 \begin{figure}[h]
 \includegraphics[width=55mm]{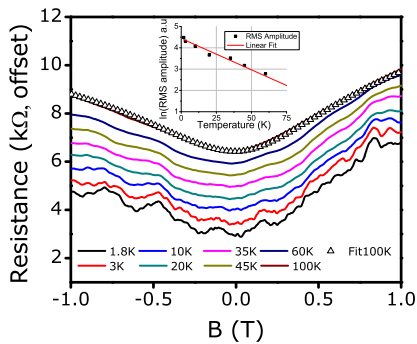}%
 \caption{(Color online) Longitudinal resistance as a function of magnetic field for different temperatures.  Oscillations are present up to 60 K.  The 1.8 K curve is at its actual value with other curves offset by 500 $\Omega$.  Inset : Linear fit to the logarithm of the RMS amplitude of the oscillations, revealing their exponential characteristic versus temperature.}
 \end{figure}

We first consider the behaviour of the conductance fluctuations away from the CNP. Figure 3(a) shows the fluctuations from a density $1.89$$\times$$10^{12}$ cm$^{-2}$ to $1.96$$\times$$10^{12}$ cm$^{-2}$ between  -1T and 1T . The data are shown as a greyscale of its field derivative. Fast Fourier Transforms (FFT) performed in this data show no significant changes if the field range is increased from 0.5T to 1T. Figure 3(b) is the equivalent plot but around the CNP. In this regime, the conductance fluctuations evolve dramatically over small changes in density. A series of parallel lines all pointing towards the CNP are observed in all four quadrants of the data at fields greater than 0.5T. The amplitude of the conductance fluctuations around the CNP is slightly higher than at higher carrier densities reaching 6\% of the total conductance compared to 2\% at $1.89$$\times$$10^{12}$ cm$^{-2}$  .

 \begin{figure}[h]
 \includegraphics[width=55mm]{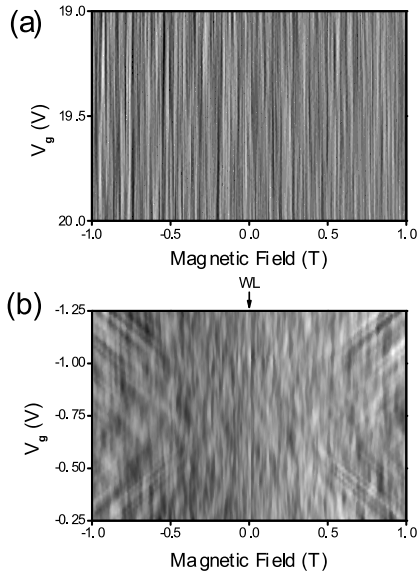}%
 \caption{a. Numerical derivative versus magnetic field  $\frac{dR_{xx}}{dB}$ for different density values in the electron regime.  Black is minimal $\frac{dR_{xx}}{dB}$ while white is maximal $\frac{dR_{xx}}{dB}$  b. Similar plot, at densities around the CNP.  CNP is at -0.75 V.}
 \end{figure}

	In Figure 4(a), we plot the raw data minus a background, which removes the dip feature (n.b. the same background is removed for all data). It is instructive to look at the FFTs as one moves through the CNP. The FFT plots, as seen Figure 4(b), (where the data are restricted to $\pm$ 0.5 T to avoid the regions with diagonal lines) show that the spread in frequencies is narrowest at the CNP and increases in either direction as the gate voltage is tuned away from the CNP. Such behaviour is evidence that the fluctuations reflect the size of the puddles in the co-existence regime. At the CNP one might expect the size of electron and hole puddles to be relatively similar. As one moves away from the CNP towards more positive (negative) voltages the size of electron (hole) puddles grows (shrinks) leading to a wider spread of frequencies. Within this picture the dominant peak at the CNP in the FFT provides a measure of the spatial extent of the underlying potential fluctuations. In our measurements, this suggests a value of 125 nm for the puddle size. While our results are totally consistent with such a scenario, we caution that the underlying magnetic field period is only slightly bigger than a quarter of the field range. The size we calculate compares with estimates from 30 to 150 nm from scanning probe experiments \cite{Martingraphenepuddles, westervelt}.

 \begin{figure}[h]
 \includegraphics[width=55mm]{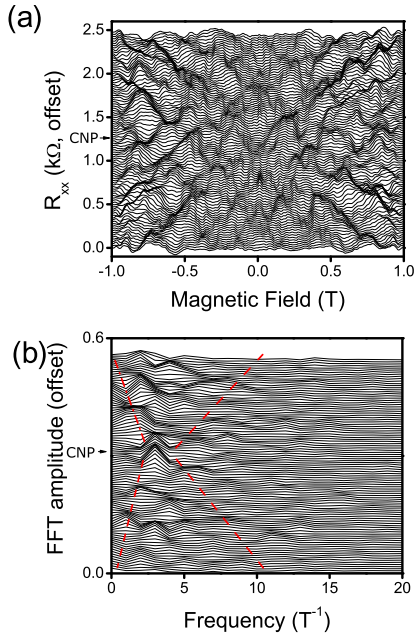}%
 \caption{(Color online) a. Waterfall diagram of fluctuations in R$_{xx}$ with background removed around the CNP.  b.  Fourier transforms of different lines in the waterfall diagram restricted to $\pm$0.5T (see text or details).  Red dashed lines are a guide to the eye emphasising the narrowing of the fluctuation range at the CNP.}
 \end{figure}
	To elucidate the nature of the parallel lines we plot, in Figure 5, the magneto-conductance over wider magnetic field and density ranges effectively expanding the measurements of the bottom right quadrant of Figure 3(b). A careful examination of the magnetic field derivative of this plot reveals two additional much weaker sequences of parallel lines. Also shown on the plot are calculated lines corresponding to filling factors 2, 6 and 10 of the quantum Hall effect. It is clear that the slopes of the three sets of parallel lines match up exactly with these filling factors. In the inset of Figure 5, we show a higher resolution resistance measurement (i.e. not the derivative) taken in one region of the plot. The full number of parallel features is more evident in these data.

 \begin{figure}[h]
 \includegraphics[width=55mm]{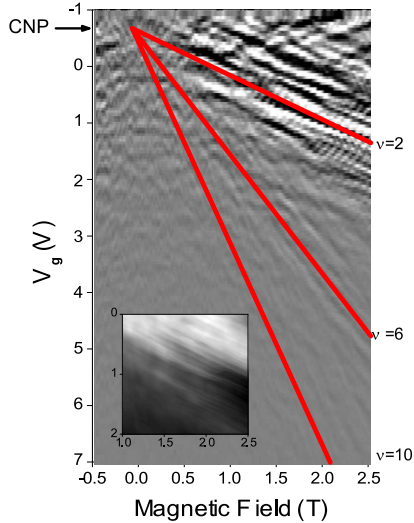}%
 \caption{(Color online) Greyscale of numerical derivatives $\frac{d^2R_{xx}}{dB^2}$, up to high electron densities.  Red lines show the theoretical positions of filling factors 2, 6 and 10 lines, from top to bottom.  Inset shows a higher resolution plot of the raw data for the upper right region.}
 \end{figure}

	We now address the origin of the lines corresponding to QHE filling factors slopes. To break down, the QHE requires edge states on opposite sides of the sample to equilibrate via backscattering events. In narrow samples this can occur resonantly via potential fluctuations. Cobden et al. \cite{cobden} showed in small MOSFETs that at high magnetic fields such resonant backscattering events followed lines parallel to the QH filling factors. They proposed a model involving interactions. The potential inhomogeneity screening ability of the 2DEG is removed as the Landau level locally approaches full occupation. This leads to small \textit{quantum dots} (QH quantum dots) isolated by incompressible regions. The backscattering events occur as the fluctuations are charged with single electrons in the same way that current flows through quantum dots at Coulomb blockade peaks. Such a charging picture has recently been confirmed by state-of the art probe experiments in AlGaAs/GaAs and Graphene 2DEGs \cite{Martin2deg, Martingraphene}. The probe experiments measured the local compressibility of these fluctuations directly. Our parallel lines originate at magnetic fields four times lower than any QHE feature and so a simple backscattering model associated with the breakdown of the QHE cannot be trivially evoked. We propose that the parallel lines are a precursor to the QHE which is possible due to the very wide density spread that occurs in graphene devices. From the spread in parallel lines we estimate density fluctuations of $2$$\times$$10^{11}$ cm$^{-2}$  for electrons consistent with that obtained from local probe techniques \cite{Martingraphene, westervelt}. This results at low fields in a complex extended state network connecting closely separated contacts at low fields. Isolated QH dots can form "locally" in this regime. At these low magnetic fields charging of these dots does not backscatter current in the sense of the resonant breakdown of the quantum Hall effect.  However, local equilibration redistributes the current in the current path network. These events are detected as fluctuations by the voltage probes. Amazingly the large density fluctuations in Graphene, $2$$\times$$10^{11}$ cm$^{-2}$,  enable this type of conductance fluctuations to be observable for 2T before even the SdH density of state oscillations are detected. Since sets of equally spread lines correspond to the charging of individual quantum dot states, we can estimate the size of the localized states. We find sets of up to 5 equally spaced lines. A typical \textit{quantum dot} size we obtain in this analysis is about 160 nm (compared to 60 nm from Martin \textit{et al.} \cite{Martingraphene} using probe techniques). With a different contact arrangement (i.e.probing a different area of the graphene flake) we observed qualitatively identical results but with a few fluctuations at twice the density spread of the results above, consistent with the mesoscopic nature of the fluctuations.	

	In conclusion, we have studied magneto-conductance fluctuations in a graphene monolayer close to the charge neutrality point. We find that the large density fluctuations in Graphene leads to two separate categories for magneto-conductance fluctuations at low fields, those related to quantum interference and those related to charging of localized quantum Hall states. The spread in the Quantum interference fluctuations in FFT plots narrows as one approach the CNP suggesting that for these fluctuations it is the size of electron/hole puddles that is important rather then the overall density fluctuation range. 

	We would like to acknowledge important motivating discussions with Louis Gaudreau, Ghislain Granger, Pawel Hawrylak, Devrim Guclu, Josh Folk and Mark Lundeberg.  A.S. and F.P. acknowledge funding from CIFAR. A.S. and S.B. acknowledge assistance from NSERC.

\bibliography{Bibliographiethese}

\end{document}